\documentclass[3p,times,twocolumn]{elsarticle}

\usepackage{ecrc}

\volume{00}

\firstpage{1}

\journalname{Nuclear Physics B Proceedings Supplement}

\runauth{}

\jid{nppp}

\jnltitlelogo{Nuclear Physics B Proceedings Supplement}

\usepackage{amssymb}

\newcommand \x {\cdot}

\newcommand \bvec{\left( \begin{array}{c} }
\newcommand \evec{\end{array} \right)}

\newcommand \bea{\begin{eqnarray} }
\newcommand \eea{\end{eqnarray} } 
\newcommand \nn {\nonumber}
\newcommand {\be} {\begin{equation}}
\newcommand {\ee} {\end{equation}}

\biboptions{compress}

\usepackage[figuresright]{rotating}

\begin{document}

\begin{frontmatter}



\dochead{}

\title{Hard Probes: After the dust has settled!}


\author{Abhijit Majumder}

\address{Department of Physics and Astronomy, Wayne State University, Detroit, Michigan 48201, USA.}

\begin{abstract}
A lot has been learnt in the 15 years since the first data on jet modification at the Relativistic Heavy Ion Collider (RHIC). These proceedings will describe the portion of the theory that is unassailable, and attempt to chart a course for the next period of precision measurements and related calculations. In particular, we will focus on the emerging series of higher order calculations, which may reveal the scale, energy and temperature dependence of jet transport coefficients, as well as develop ingredients for a future NLO calculation of jet modification. Connections with the underlying degrees of freedom in the Quark Gluon Plasma (QGP) will be explored. We will discuss the challenges faced by Monte-Carlo event generators and their comparison with a broad spectrum of jet modification data. 
First principles calculations of transport coefficients, and the determination of new transport coefficients which encapsulate the transition of energy momentum from the jet scale to the medium scale will also be discussed.
\end{abstract}

\begin{keyword}
Jet quenching, energy loss, Quark Gluon Plasma, Monte-Carlo event generator

\end{keyword}

\end{frontmatter}


\section{The continuing debate}
\label{hist}

Jet quenching was experimentally observed, in the first set of experiments at the Relativistic Heavy-Ion 
Collider (RHIC) (see Refs.~\cite{Adcox2005184,Adams:2005dq} for a review). Even before the 
experimental observation, there existed at least 3 different (and differing) theoretical approaches based on weak coupling perturbative QCD (pQCD): 
The Gyulassy-Levai-Vitev (GLV) approach~\cite{Gyulassy:1999zd,Gyulassy:2000er,Gyulassy:2000fs}, 
the Baier-Dokshitzer-Mueller-Peigne-Schiff approach~\cite{Baier:1994bd,Baier:1996kr,Baier:1996sk}, 
and the Higher-Twist approach~\cite{Wang:2001ifa,Guo:2000nz}. 
With the advent of the Arnold-Moore-Yaffe (AMY) approach~\cite{Arnold:2002ja,Arnold:2001ms,Arnold:2001ba}, 
the early years of jet quenching at RHIC had become a rather unclear 
exercise. 
Each of these formalisms engenders somewhat different physical approximations. 
In spite of these differences in physical content, all four of these 
approaches were able to describe some of the data on the suppression of high $p_T$ particles at RHIC. All of these have at least one adjustable parameter, e.g., $\hat{q}$, opacity, $\alpha_S$ etc., which may be 
tuned to a single data point. The lack of sufficient high-$p_T$ data at RHIC prevented one from discerning differences between these various approaches. 
Along with approaches based on pQCD, there also arose models based on the strong coupling AdS/CFT correspondence~\cite{Liu:2006ug,CasalderreySolana:2007qw,Herzog:2006gh,Gubser:2006bz,Arnold:2010ir,Arnold:2011qi}.

This situation changed quickly with the start of the LHC, which not only challenged the existing models 
of energy loss (in particular those based entirely on strong coupling) 
but also expanded the range of observables that had to be describable by energy loss approaches, 
from leading hadrons to full jets. This has led to a lot of development within the various approaches to jet modification. 
In particular, the focus on full jets, has resulted in a movement within energy loss formalisms towards Monte-Carlo event generators~\cite{Schenke:2009gb,Renk:2010zx,Zapp:2008gi,Zapp:2012ak,Majumder:2013re,Majumder:2014gda,He:2015pra}.

While these generators have greatly improved the 
functionality of jet modification approaches, allowing them to address a wide variety of jet observables, 
they have not yet addressed the veracity of the underlying physical picture, and in many cases have introduced additional uncertainties.
These proceedings will highlight these issues and attempt to outline the form of a final event generator that will have to be constructed 
to resolve the continuing disagreement between the various energy loss models.

\section{A schematic patchwork}

A high energy jet is formed in a very hard interaction between two incoming partons in two opposing nucleons. A hard parton produced in such an interaction is typically far off its mass shell in a high virtuality state. As it proceeds through the medium it will radiate numerous gluons and lose virtuality as it 
propagates through the medium. 
The interaction of a parton within the jet with the deconfined medium is dictated by the energy and virtuality of the hard parton. In the following we outline, using handwaving arguments, the evolution of the jet and the interaction with the medium through which it propagates. 

Let us consider a parton produced in the final state with an energy $E$ and virtuality $\mu$, such that  $E \gg \mu \gg \Lambda_{\rm QCD}$. 
Up to an order of magnitude, the lifetime and transverse size of such a parton is given as, 
\bea
\tau \sim E/ \mu^2; \,\,\,\,\,\,\,\,\,\,\, \delta x^2 \sim 1/k_\perp^2 \sim 1/\mu^2 \ll 1/T^2,  \label{LifetimeSize}
\eea
where, $k_\perp$ is the transverse momentum of either daughter particle with respect to the axis of momentum of the parent, and $T$ is the 
temperature of the medium and a very approximate estimate of the size of a degree of freedom in the plasma. As a result, highly virtual partons 
within the jet are much smaller than the transverse size of a plasma degree of freedom. As a result, the virtual parton will scatter off a parton within 
the plasma degree of freedom. The parton distribution function that is sampled by the hard parton, depends on the energy of the parton, or alternatively 
on the ``$x$'' (momentum fraction) of the distribution, as well as the virtuality or transverse size of the hard parton. In deep-inelastic scattering, 
the momentum fraction is defined as, 
\bea
x &=& Q^2/(2 E M),
\eea
where, $M$ is the mass of the nucleon, and $E$ is effectively the energy of the hard photon. Since the mass of a plasma degree of freedom, or a color correlated 
regime of the plasma is unknown, $x$ dependence is understood as an inverse dependence on $E$.

Within this set up, 
the partons within a jet fall within three regimes which we discuss briefly below:

1) The high-energy-high-virtuality regime: This phase describes partons within the jet from production to the point where either the Energy $E$ becomes 
comparable to $\Lambda_{QCD}$, or the virtuality of the parton has approaches the saturation scale $Q^2 \sim \hat{q} \tau$, where $\tau$ is the lifetime of the
 hard parton (see above). In this phase, the splittings of the partons are predominantly determined by the large off-shellness of the partons; scattering with the medium tends to modify this off-shellness and the ensuing development of the shower. An alternative means of describing this phase is that radiations (from splits) dominate over scatterings of the partons. Scattering and emission are described using the high-$Q^2$ portion of the higher twist (HT) energy loss scheme. 
 
 In this regime, multiple emissions are calculated by using Sudakov form factors based on the Dokshitzer-Gribov-Lipatov-Altarelli-Parisi (DGLAP) emission spectrum. Radiations are strongly ordered in transverse momentum (or angle)  and thus interference between multiple hard emissions are suppressed. This represents a phase of vacuum like radiation, where the radiation 
 rate is perturbatively influenced by scattering in the medium. Partons that are considerably more virtual than the temperature scale  of the medium will sample 
 short distance structure within the QGP degrees of freedom. This sampled distribution of partons, similar to the parton distribution function of a nucleon will 
 evolve with scale and energy of the hard parton.  Currently this phase is described by the YaJEM~\cite{Renk:2010zx} and MATTER~\cite{Majumder:2013re} event generators. 
 \begin{figure}[h!]
\begin{center}
\vspace{-0.35cm}
\resizebox{3in}{3in}{\includegraphics{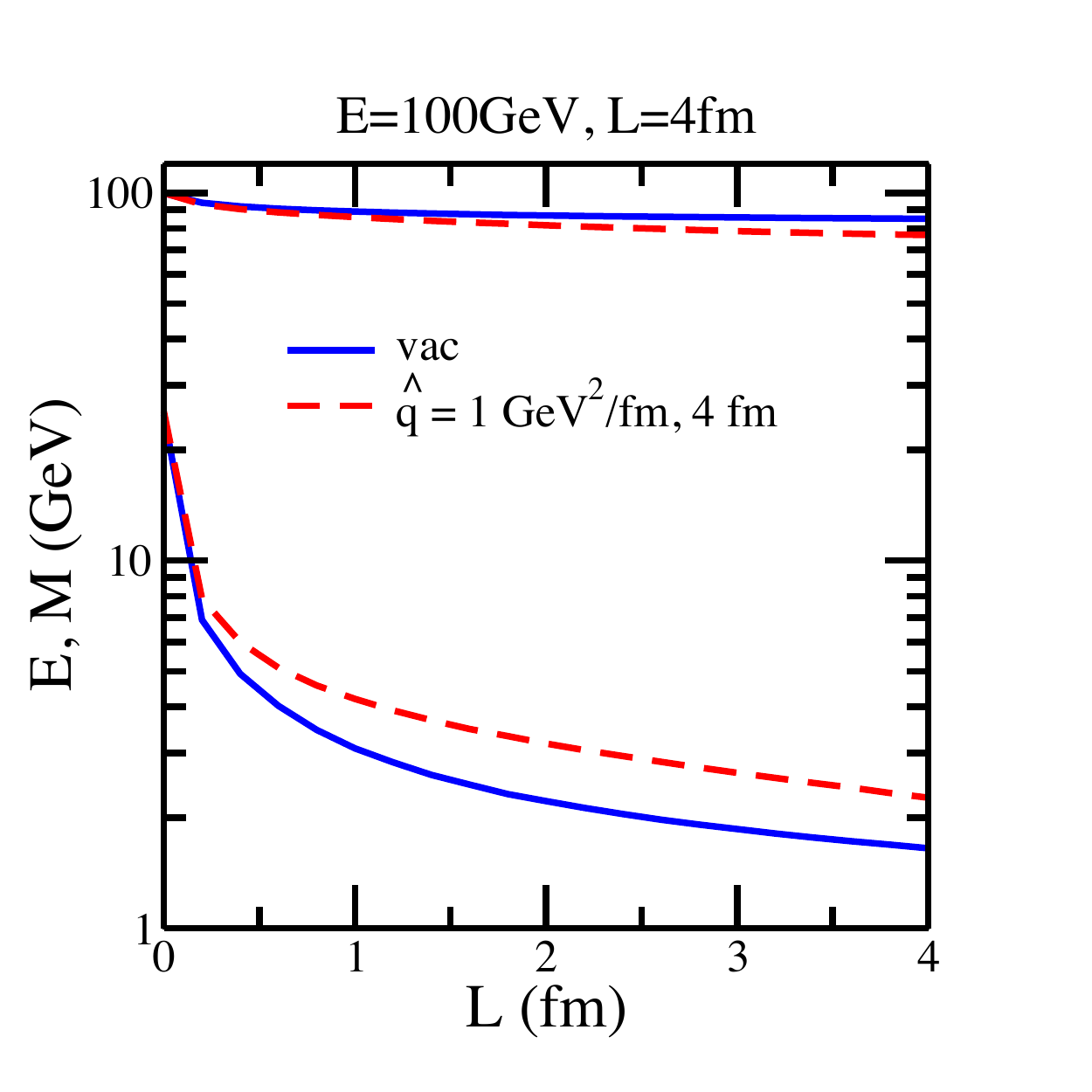}} 
\vspace{-0.75cm}
    \caption{Energy and virtuality, as a function of length, of the leading parton of a $100\,$GeV jet traveling through a 4\,fm long dense medium, held at a fixed temperature. See text for details and discussion.}
    \label{LengthEnergyVirtuality}
 \end{center}
\end{figure}
 
 2) The high-energy-low-virtuality regime: With sequential emissions in the HT-DGLAP phase, the virtuality of the partons in the shower continues to drop with each emission. Perturbative branchings tend to be highly asymmetric and as such the leading partons in the shower carry a large fraction of the energy of the jet. In spite of the large energy within the leading partons, the virtuality tends to drop swiftly as the partons traverse the plasma. 
 Plotted in Fig.~\ref{LengthEnergyVirtuality} is the energy and virtuality of the leading parton in a 100 GeV jet produced at one end of a 4 fm long container that is filled with a QGP at a fixed temperature. It is assumed that this medium will produce an effective $\hat{q} = 1$~GeV$^2$/fm. 
 The figure includes the energy and virtuality of the parton passing through the QGP and a similar parton in vacuum. It is clear from the figure that 
 the leading parton loses virtuality much more rapidly than energy. Also clear, is the fact that partons in the medium lose virtuality somewhat more slowly than partons in vacuum. This is due to scattering in the medium, which tends to hamper the drop in virtuality. Beyond this length,  
 the few scattering approximation is no longer valid. Partons in this regime are in the many scatterings or BDMPS regime. 
 
 For partons in this regime, the mean virtuality scales with energy of the parton. Note:  $Q^2 \sim \hat{q} \tau$ and as a result, 
 \bea
 Q^2  &\sim& \sqrt{ \hat{q} E  }. 
 \eea
 In this regime, the hard parton engenders multiple scattering between emissions. 
 Multiple scattering over time tends to increase the off-shellness of the parton, which drops with the eventual emission.
 Such hard emissions are rarer, suppressed by $\alpha_S$. Successive emissions are not ordered in virtuality or transverse momentum. However, the 
 large separation in time ensures that the emissions are independent. The evolution of the jet in such a scenario is calculated using a rate equation. 
 The medium, in this phase, is sampled at the scale of $\hat{q}\tau$, and as a 
 result, the resolution of the medium changes much more slowly per scattering as compared to the high energy high virtuality phase. 
Currently this phase is described by the AMY formalism for event averaged leading hadrons and by the JEWEL~\cite{Zapp:2008gi,Zapp:2012ak} 
and MARTINI~\cite{Schenke:2009gb} Monte-Carlo generators for the event-by-event observables. 

At the time of writing, all calculations in the high-energy-low-virtuality phase (via AMY, MARTINI or JEWEL) ignore the existence of the earlier higher virtuality phase. 
Similarly all calculations in the high virtuality phase ignore the existence of the later lower virtuality phase. 
The fate of the radiation spectrum in the region intermediate between these two regimes is also unknown. 
To model this transition regime, one needs to consider radiation at NLO (currently being carried out~\cite{Fickinger:2013xwa}) and perhaps even at NNLO, to understand the systematics of how one transitions from a transverse momentum (or angular) ordered shower to a shower with time separated emissions.
In the absence of such calculations, no mechanism is known that allows one to numerically simulate this region. This remains a major area of research for future generators.

 3) The low-energy-low-virtuality regime: Quickly after the initial hard emissions, the partons in a shower, on average, organize into a pattern of a few, very high energy partons, and several soft partons. The high energy partons remain weakly coupled with the medium due to the higher virtuality generated by multiple scattering. The soft partons, whose energy and virtuality has dropped down to a few GeV will be strongly coupled with the medium. The only known means to describe this phase of the shower is by using the techniques of the AdS/CFT conjecture~\cite{Gubser:2006bz,Arnold:2010ir,Arnold:2011qi}. Such partons are continuously emitted by the harder partons in the jet and are expected to thermalize swiftly within the medium. 
 
With the exception of one hybrid formalism~\cite{Casalderrey-Solana:2014bpa}, which considers a vacuum shower undergoing perturbative splits, unmodified by the medium, with each parton losing energy via strong-coupling dynamics, there has also been no attempt to incorporate strong and weak coupling energy loss approaches within a single formalism. Prior work which had entirely ignored weak coupling energy loss in favor of a pure strong coupling approach have been ruled out 
by experimental results. 
Beyond strong coupling approaches, several other, predominantly weak coupling based event generators have tended to introduce phenomenological corrections to deal with this strongly-coupled phase.

 \section{From Leading Hadrons to Full Jets}

 Beyond the three regimes mentioned above, there are several other components to the object that is collected as a jet in the detectors. Experimentally, there is no means to differentiate between the hadrons that emanate from the medium and those from the fragmentation of partons within the jet that have escaped the medium. 
Several statistical subtraction routines have been attempted to extract the jet from the underlying medium. These include both a subtraction of the expected mean energy of the medium from a jet identified with an anti-$k_{T}$ reconstruction routine, and an unfolding of the spectrum to remove the effect of the fluctuating medium (In some cases, constituent cuts have also been attempted). Such reconstruction is complicated by the fact that energy lost by a jet fluctuates from event to event even in a static medium. In a fluctuating medium, the mean energy and fluctuation of energy lost by the jet are both effected by the medium. As jets lose energy-momentum in a medium, this energy will tend to thermalize, with some portion remaining within the jet cone and some escaping outward. Along with these non-perturbative mechanisms, there could also exist genuine perturbative mechanisms that transport energy out of the jet cone~\cite{Fister:2014zxa,Blaizot:2014ula,Blaizot:2013hx}.
The entire dynamics of energy loss and partial thermalization in the QGP is always followed by the process of hadronization. While the hadronization of the soft sector of the QGP, in the absence of jets, is under some measure of phenomenological control, there is wide uncertainty in the hadronization (fragmentation) of the hard portion of the jet in the vicinity of the dense medium. So far this is modeled using some combination of string fragmentation and recombination approaches. These approaches also have an effect on what portion of the jet ends up in the reconstructed cone.

\begin{figure}[h!]
\resizebox{3in}{4in}{\includegraphics{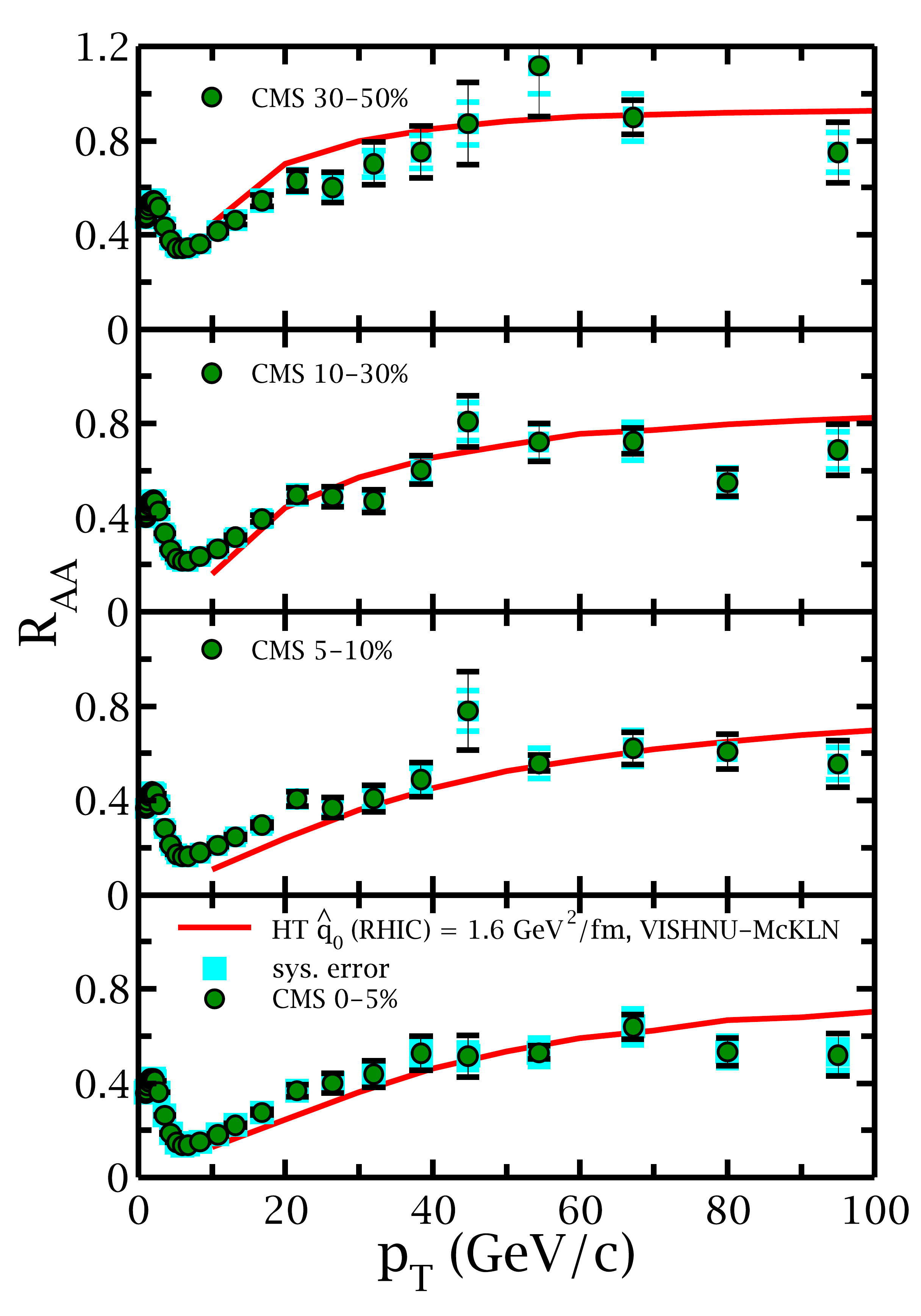}} 
\vspace{-0.5cm}
    \caption{Theoretical calculations, within the Higher-Twist scheme, of the Nuclear Modification factor at four different centralities of collision, compared with measurements by the CMS detector at the LHC.}
    \label{RAA}
\end{figure}

 In spite of the lack of a complete formalism for jet energy loss, there has been quite some success by event averaged calculations of leading hadron suppression. This is primarily due to the fact that leading hadrons emanate from the highest energy portion of the jet, and in most cases arise from fragmentation in vacuum. The highest energy portion of the jet depends solely on only two portions of the physical process described above: the high-virtuality-high-energy regime and the high-energy-low-virtuality regime.  Calculations of energy loss using either of these formalisms has shown the ability to describe the data on the nuclear modification factor $R_{AA}$ as a function of $p_T$ and centrality, 
 as shown in Fig.~\ref{RAA}. 
Similar results have also been reported by the AMY scheme. 
The precise reason for this success by two different approaches is not completely understood, but probably has to do with the inclusive nature of the observable and the possible theoretical overlap between the two 
approaches~\cite{Majumder:2010qh,Majumder:2007iu}.

Contrary to popular belief, these leading hadron observables are currently, quite sufficient, to allow for the determination of the traditional transport coefficients, such as the transverse momentum diffusion parameter 
$\hat{q}$~\cite{Baier:2002tc} and the drag coefficient $\hat{e}$~\cite{Majumder:2008zg}, defined as,
\bea
\hat{q} = \frac{\langle p_\perp^2 \rangle}{L},  \,\,\,\,\,\,\,\,\,\,\,\,\, \hat{e} = \frac{\langle p_z \rangle}{L},
\eea
where, $\langle p_\perp \rangle$ and $\langle p_z \rangle$ represent the transverse and longitudinal momentum exchanged between a single parton and 
the dense medium. Since these partons are close to mass shell, one often replaces $p_z$  with the energy $E$ or even the light-cone momentum.
 
Full jets, necessarily, also involve other new coefficients that are yet to be determined. As a result, while full jet analysis does indeed have the potential to further constrain the standard transport coefficients, observables need to be devised that are insensitive to the new transport coefficients.

\section{Transport coefficients: new and old}

The information extracted from the modification of hard jets is expressed in the form of transport coefficients. To consider the modification 
of full jets, one necessarily has to consider two different types of transport coefficients, which we denote as type I and type II. Type I coefficients, 
like $\hat{q}$ and $\hat{e}$, defined above, describe simply the transfer of energy between the partons and the medium. 
Such coefficients are not simply limited to only $\hat{q}$  and $\hat{e}$. There is a distribution of the exchanged momentum between the hard parton and the soft medium and $\hat{q}$ represents the second moment of this distribution (the first moment is assumed to be vanishingly small), and has the field theoretic definition
\bea
\hat{q} (x) &=& \frac{4 \pi^2 \alpha_S}{N_C} \int d y^- d^2 y_\perp \frac{d^2 k_\perp}{(2 \pi)^2}  e^{- i xP^+  y^-  + k_\perp \cdot y_\perp  }  \nn \\
&\times & \langle n | \frac{e^{-\beta E_n}}{ Z } \mathcal{U}^\dag F^{+ \mu }  (y^-, y_\perp) \mathcal{U} F^+_\mu (0,0)  | n \rangle.
\eea
The equation above has been derived for a single parton that scatters once off a dense QCD medium. The $F^{+ \mu} = t^a {F^a}^{+ \mu }$ represent the chromo-magnetic field strength operator. The state $|n \rangle$ represents one state with energy $E_n$ in an enclosed static medium, in contact with a heat reservoir at temperature $1/\beta$. 
In the equation above, $k_\perp$ is the momentum transverse to the direction of a single hard parton exchanged between the parton and the medium.
The value of $x P^+$  represents the light-cone momentum exchanged between the parton and the medium, where $P^+ $, in the rest frame of the bath, is the mean mass of a correlated  section of the medium. The $\mathcal{U}$ represents a Wilson line in the adjoint representation, introduced to make the expression gauge invariant (trace over the operator product is implied).

As can be seen from the equation above, $\hat{q}$ is a function of the light-cone momentum fraction $x$, and if calculated at next-to-leading order also of the factorization scale $\mu$. This dependence is beyond the dependence of $\hat{q}$ on the temperature of the medium. 
At the time of writing there exist several higher order calculations of $\hat{q}$, in different regimes: one in the high $Q^2$ regime~\cite{Kang:2013raa} 
and several in the high-$E$ and low-$Q^2$ regime (or low-$x$ regime)~\cite{Abir:2015qva,Blaizot:2014bia,Iancu:2014kga,Liou:2013qya,Ghiglieri:2015ala}, which yield different results, due to the different set of contributions that are being resumed at NLO, in the different regimes. 

Beyond $\hat{q}$ and $\hat{e}$ one may define a series of transport coefficients that involve higher moments of the momentum exchanged between the hard jet and the medium:
\bea
\hat{q}_4 &=& \frac{\langle p_\perp^4 \rangle - \langle p_\perp^2 \rangle^2}{L}, \cdots \nn \\
\hat{e}_2 &=& \frac{\langle \delta p_z^2 \rangle }{  L} , \,\,\,\,  \hat{e}_4 = \frac{\langle \delta p_z^4 \rangle - \langle \delta p_z^2 \rangle^2}{L}, {\rm etc.}
\eea
These transport coefficients describe the energy-momentum exchange between the hard parton and the medium. They contain
no information about the fate of the energy-momentum that is deposited in the medium from the hard parton, i.e., if this energy thermalizes with the 
medium and leaves the jet cone, or if some portion of it remains within the jet cone. Another issue is that these are all hard coefficients, i.e., they involve a hard parton that interacts with a soft non-perturbative medium which is factorized from the hard parton. 

To understand the physics of energy deposition and thermalization, one needs to look at hard-soft coefficients, where the energy deposited in a small transverse space around the hard jet grows from the inverse jet scale $1/Q^2$ to the thermal scale $1/T^2$ and beyond, when it becomes an energy-momentum source term in the medium. 
As an illustration, we propose a representative form of the transverse shape of the energy density source term: 
\bea
\langle \phi (x,y,t) \rangle = \frac{N}{4\pi \sigma(T,Q) t} \exp \left[ - \frac{x^2 + y^2}{4 \sigma(T,Q) t} \right]. 
\eea  
 The equation above is meant to describe the two dimensional Gaussian diffusion of energy deposited at a location $(x,y,z)=0$ as a function of $t$, the time since the hard parton, traveling in the z direction, has passed through this location. The new transport coefficient is $\sigma(T,Q)$ which depends on both the temperature and the virtuality of the parton that has deposited the energy. So far no such transport coefficients have been considered in the redistribution of the energy deposited by a shower, within a jet cone. However, there have been several attempts to model this redistribution~\cite{Qin:2009uh,Renk:2013pua,Neufeld:2009ep,Neufeld:2011yh}
The development of a formalism of full jet analysis that includes both types of transport coefficients remains an outstanding problem in energy loss.

\section{Outlook and Acknowledgements}

Several of the issues mentioned above were discussed at the Hard Probes 2015 conference. Several improvements in the formulation and implementation of different approaches to energy loss in dense matter were  highlighted. However, there remains a lack in attempts to incorporate different approaches into a single overarching formalism. This proceeding, as well as the talk on which it was based, were written in an attempt to seed such a discussion in future efforts. 

This work was supported in part by the US National Science Foundation under grant number PHY-1207918 and by the US Department of Energy under grant number DE-SC00013460. 



\bibliographystyle{elsarticle-num}
\bibliography{majumder_refs}







\end{document}